\documentstyle[12pt]{article}

\def\+{{+\!\!\!+}}
\def\pp{\mbox{\tiny${}_{\stackrel\+ =}$}}

\def\d{\partial}

\def\th{\theta}

\def\P{\Phi}

\def\p{\psi}

\def\e{\varepsilon}

\def\bR{\hbox{R\hspace{-0.10in}I\hspace{0.04in}}} 
\def\pmb#1{\setbox0=\hbox{#1}%
\kern.0em\copy0\kern-\wd0
\kern-.04em\copy0\kern-\wd0
\kern.08em\copy0\kern-\wd0
\kern-.04em\raise.0433em\box0 }         

\def\half{\frac{1}{2}}
\def\rank{\textstyle{\rm{rank}}}
\def\diag{\textstyle{\rm{diag}}}
\newcommand{\nc}{\newcommand}
\nc{\beq}{\begin{equation}}
\nc{\eeq}[1]{\label{#1}\end{equation}}
\nc{\ber}{\begin{eqnarray}}
\nc{\eer}[1]{\label{#1}\end{eqnarray}}
\nc{\pek}[1]{\cite{#1}}
\nc{\enr}[1]{(\ref{#1})}
\nc{\kal}[1]{{\cal{#1}}}
\nc{\dott}{\;\cdot\;}

\begin{document}
\newcommand{\inv}[1]{{#1}^{-1}} 
\renewcommand{\theequation}{\thesection.\arabic{equation}}
\newcommand{\be}{\begin{equation}}
\newcommand{\ee}{\end{equation}}
\newcommand{\bea}{\begin{eqnarray}}
\newcommand{\eea}{\end{eqnarray}}
\newcommand{\re}[1]{(\ref{#1})}
\newcommand{\qv}{\quad ,}
\newcommand{\qp}{\quad .}
\begin{center}
                       
                                \hfill   hep-th/0111161\\
\vskip .3in \noindent

\vskip .1in

{\large \bf{N=1 supersymmetric sigma model with boundaries, I}}
\vskip .2in

{\bf Cecilia Albertsson}$^a$\footnote{e-mail address: cecilia@physto.se},
{\bf Ulf Lindstr\"om}$^a$\footnote{e-mail address: ul@physto.se}
 and  {\bf Maxim Zabzine}$^{b}$\footnote{e-mail address: zabzine@fi.infn.it} \\


\vskip .15in

\vskip .15in
$^a${\em  Institute of Theoretical Physics \\
Stockholm Centre for Physics, Astronomy and Biotechnology \\
University of Stockholm \\
SE-106 91 Stockholm, Sweden}\\
\vskip .15in
$^b${\em INFN Sezione di Firenze, Dipartimento di Fisica\\
 Via Sansone 1, I-50019 Sesto F.no (FI), Italy}

\bigskip


 \vskip .1in
\end{center}
\vskip .4in

\begin{center} {\bf ABSTRACT } 
\end{center}
\begin{quotation}\noindent 
  We study an $N=1$ two-dimensional non-linear sigma model with
  boundaries representing, e.g., a gauge fixed open string.  We
  describe the full set of boundary conditions compatible with $N=1$
  superconformal symmetry.  The problem is analyzed in two different
  ways: by studying requirements for invariance of the action, and by
  studying the conserved supercurrent.  We present the target space
  interpretation of these results, and identify the appearance of
  partially integrable almost product structures.
\end{quotation}
\vfill
\eject

\section{Introduction}

The two-dimensional non-linear sigma model with boundaries and $N=1$
worldsheet supersymmetry plays a prominent role in the description of
open Neveu-Schwarz-Ramond (NSR) strings.  Studying this model may thus
help to understand those aspects of D-brane physics where the sigma
model description is applicable. The model is also interesting in its
own right; for instance, it is a well known fact that supersymmetric
models have intriguing relations with geometry. Here we will
show that the supersymmetric sigma model with boundaries naturally
leads to the appearance of partially integrable almost product
geometry.
 
The idea is to look at the non-linear sigma model with the minimal
possible amount of worldsheet supersymmetry, i.e.~with only one spinor
parameter in the supersymmetry transformations. It turns out that even
this minimal amount of supersymmetry leads to interesting
restrictions on the allowed boundary conditions.  These worldsheet
restrictions can be reinterpreted in terms of the target space
manifold, with the result that the open strings are allowed to end
only on (pseudo) Riemannian submanifolds of the target space.

Special cases of the kind of conditions we present here are often
adopted in the literature (e.g.~\cite{Ooguri:1996ck,Alvarez}), without
derivation. The aim of this paper is to show, in a pedagogical manner,
how one arrives at these conditions via systematic analysis of the
action and the conserved currents. However, the conditions we find are
more general than the ones usually assumed.

The paper is organised as follows. In Section~\ref{s:action} we
consider the non-linear sigma model action defined on a (pseudo)
Riemannian target space manifold. We derive the general boundary
conditions that simultaneously set to zero the field variation and the
supersymmetry variation of the action. We check that these conditions
are compatible with the supersymmetry algebra, and then consider a
special case. In Section~\ref{s:currents} we rederive the boundary
conditions using a different approach, namely requiring $N=1$
superconformal invariance at the level of the conserved currents.  In
Section~\ref{s:geometric} we interpret the worldsheet conditions in
terms of the target space manifold, which leads to some interesting
properties of D-branes.  In Section~\ref{s:moresusy} we briefly touch on 
 the interesting question of additional supersymmetries, and
 finally, in Section~\ref{s:discussion}, we 
give a summary of the present work and an outline of our plans for
future investigations.

\section{The ${\bf N=1}$ ${\bf \sigma}$-model action}
\label{s:action}

The first approach we take in finding the boundary conditions is to
analyze the action. Field variations of the action yield boundary
field equations, which must be compatible with the vanishing of the
supersymmetry variation of the action.  Making a general (linear)
ansatz for the relation between worldsheet fermions on the boundary,
this compatibility requirement imposes restrictions on our ansatz.

\subsection{Boundary equations of motion}

We start from the non-linear sigma model action 
\beq
S=  \int d^2\xi d^2\theta\,\, D_{+} \Phi^\mu D_{-}
\Phi^\nu g_{\mu\nu} (\Phi),
\eeq{acrbas}
where $g_{\mu\nu}$ is a Riemannian\footnote{Hereafter,
  whenever we use the word ``Riemannian,'' we mean ``Riemannian or
  pseudo Riemannian.''} metric and we use the standard 2D superfield
notation (see Appendix~\ref{a:11susy}).  Varying $S$ with respect to
the fields $\Phi^\mu$ we obtain a boundary term
\ber
\nonumber \delta S &=& i \int d\tau \left [ (\delta\psi_{+}^\mu \psi_+^\nu -
  \delta \psi_{-}^\mu \psi_-^\nu)g_{\mu\nu} + \right. \\
&& \left. + \delta X^\mu ( i
  g_{\mu\nu}(\partial_\+ X^\nu - \partial_= X^\nu) +
  (\psi_-^\nu\psi_-^\rho - \psi_+^\nu \psi_+^\rho
  )\Gamma_{\nu\mu\rho})\right]_{\sigma=0,\pi} \,\,\, ,
\eer{fullvar}
where $\Gamma_{\nu\mu\rho}$ is the Levi-Civita connection,
\beq
\Gamma_{\nu\rho\mu}=g_{\nu\sigma} \Gamma^{\sigma}_{\,\,\rho\mu}
 =\frac{1}{2}( g_{\nu\mu,\rho} +g_{\rho\nu,\mu} - g_{\rho\mu,\nu}) .
\eeq{Gammapm}

We now make a general linear ansatz for the fermionic boundary
conditions\footnote{From
  now on we shall drop the notation $\vert_{\sigma=0,\pi}$, since all
  conditions in this paper are understood to hold on the boundary.}
(see \cite{BorlafLozano} for a discussion of general fermionic
ans\"atze),
\beq
 \psi_-^\mu = \eta R^\mu_{\,\,\nu} (X) \,\psi_{+}^\nu \,\,\,
\vert_{\sigma=0,\pi} ,
\eeq{fermbc}
 where $R^\mu_{\,\,\nu}$ is a $(1,1)$ ``tensor'' which squares to one,
\beq
  R^\mu_{\,\,\nu} R^\nu_{\,\,\rho} = \delta^\mu_{\,\,\rho},
\eeq{RRI}
and we have found it convenient to introduce the parameter $\eta$
which takes on the values $\pm 1$. The property (\ref{RRI}) can be
justified by worldsheet parity, i.e., the theory should be invariant
under interchange of $\psi_{+}$ and $\psi_{-}$.  Relaxation of this
property leads to the appearance of generalised torsion (i.e., the
combination $B+dA$ of a $B$-field and gauge field); however, we focus
here on a torsion-free background.\footnote{The general case will be
  treated in a forthcoming publication \cite{now} (see also
  \cite{Haggi-Mani:2000uc}).}

We view $R^\mu_{\,\,\nu}$ as a formal object, for the moment avoiding
to specify whether it is defined globally or only locally; such issues
are discussed in Section~\ref{s:geometric}.  Substituting the ansatz
(\ref{fermbc}), as well as its field variation, into (\ref{fullvar})
yields conditions on $R^\mu_{\,\,\nu}$ if the variation $\delta S$ is
to vanish. The first such condition comes from cancelling the $\delta
\psi_{+}^\mu$-terms and says that $R^\mu_{\,\,\nu}$ must preserve the
metric,
\beq
 g_{\mu\nu} =  R^\sigma_{\,\,\,\mu}  R^\rho_{\,\,\,\nu} g_{\sigma\rho} .
\eeq{RRcond}
Note that this condition, together with (\ref{RRI}), implies that
$R_{\mu\nu}\equiv g_{\mu\rho}R^\rho_{\,\,\,\nu}$ is symmetric.
After imposing (\ref{RRcond}), we are left with the $\delta X^\mu$-part
of (\ref{fullvar}),
\beq
\delta S = i \int d\tau\,\, \delta X^\mu \left [ i
  g_{\mu\nu}(\partial_\+ X^\nu - \partial_= X^\nu) +
  R^\nu_{\,\,\sigma} g_{\nu\rho} \nabla_\mu R^\rho_{\,\,\gamma}
  \psi_{+}^\sigma \psi_{+}^\gamma \right ],
\eeq{fullvar1}
where $\nabla_\mu$ is the spacetime Levi-Civita covariant derivative.

Before exploring what conditions the vanishing of (\ref{fullvar1})
gives us, we first remark on the interpretation of $R^\mu_{\,\,\nu}$.
The open string may either move about freely, in which case its ends
obey Neumann boundary conditions; alternatively, the ends of the
string may be confined to a subspace, corresponding to Dirichlet
conditions.  Consider a $d$-dimensional target space with Dp-branes,
i.e., there are $d-(p+1)$ Dirichlet directions along which the field
$X^\mu$ is frozen ($\partial_0 X^i =0, i=p+1,...,d-1$). At any given
point on a Dp-brane we can choose local coordinates such that $X^i$
are the directions normal to the brane and $X^m$ ($m=0,...,p$) are
coordinates on the brane. We call such a coordinate system
\emph{adapted} to the brane. In this basis the fermionic boundary
conditions are \cite{Polchinski96}
$$
\psi_-^m = \eta \,\psi_{+}^m ,
\,\,\,\,\,\,\,\,\,\,\,\,
\psi_-^i = -\eta \,\psi_{+}^i ,
$$
whence
\beq
 R^\mu_{\,\,\,\nu} = \left ( \begin{array}{cc}
            \delta^n_{\,\,\,m} & 0 \\
              0 & -\delta^i_{\,\,\,j}
           \end{array} \right ) .
\eeq{bldfR}
(This tensor is also used in boundary state formalism, see
e.g.~\cite{Hashimoto,Fotopoulos}.)  So $R^\mu_{\,\,\,\nu}$ has a clear
physical interpretation: its ($+1$) eigenvalues correspond to Neumann
conditions and its ($-1$) eigenvalues correspond to Dirichlet
conditions.  Thus the general tensor $R^\mu_{\,\,\nu}$ represents the
boundary conditions covariantly at the given point.

Given this setup of Neumann and Dirichlet directions, it is
convenient to introduce the objects $P^\mu_{\,\,\nu}$ and
$Q^\mu_{\,\,\nu}$ as \cite{Fotopoulos}
\beq
P^\mu_{\,\,\nu} =\frac{1}{2}(\delta^\mu_{\,\,\nu} +
R^\mu_{\,\,\nu}),
\,\,\,\,\,\,\,\,\,\,\,
Q^\mu_{\,\,\nu}
=\frac{1}{2}(\delta^\mu_{\,\,\nu} - R^\mu_{\,\,\nu}).
\eeq{PQdef}
Because $R$ squares to one, $P^\mu_{\,\,\nu}$ and
$Q^\mu_{\,\,\nu}$ are orthogonal projectors, defining the Neumann and
Dirichlet directions, respectively, in a covariant way.  They satisfy
\beq
P^\mu_{\,\,\rho}P^\rho_{\,\,\nu} = P^\mu_{\,\,\nu},
\,\,\,\,\,\,\,\,\,\,\,
Q^\mu_{\,\,\rho}Q^\rho_{\,\,\nu} = Q^\mu_{\,\,\nu},
\eeq{e:PQrel1}
and
\beq
Q^\mu_{\,\,\nu}P^\nu_{\,\,\rho}=P^\mu_{\,\,\nu}Q^\nu_{\,\,\rho}=0,
\,\,\,\,\,\,\,\,\,\,\,
P^\mu_{\,\,\nu}+Q^\mu_{\,\,\nu} =\delta^\mu_{\,\,\nu}.
\eeq{e:PQrel2}

Continuing the analysis of the field variation
(\ref{fullvar1}), we note
that since, on the boundary, the $X$-field is frozen along
the Dirichlet directions, its corresponding variation vanishes.
That is, $Q^\mu_{\,\,\nu}\delta X^\nu =0$,
and we may write $\delta X^\mu = P^\mu_{\,\,\nu}\delta X^\nu$.
Thus the
set of general parity invariant boundary conditions implied by the
boundary equations of motion are
\beq
\left \{ \begin{array}{l}
 \psi_{-}^\mu - \eta R^\mu_{\,\,\nu} \psi_{+}^\nu = 0 \\
 g_{\delta\mu} P^\mu_{\,\,\nu}( \partial_{\+}X^\nu - \partial_{=} X^\nu)
 - i P^\mu_{\,\,\delta} R^\nu_{\,\,\sigma}g_{\nu\rho} \nabla_\mu
 R^\rho_{\,\,\gamma} \psi_{+}^\sigma \psi_+^\gamma =0 \\
 Q^\mu_{\,\,\nu}(\partial_{=}X^\nu + \partial_{\+} X^\nu) =0
\end{array} \right .
\eeq{genfullNbc1}
where $R^\mu_{\,\,\nu}$ satisfies
\beq
\left \{ \begin{array}{l}
 R^\mu_{\,\,\nu} R^\nu_{\,\,\rho} = \delta^\mu_{\,\,\rho} \\
 g_{\mu\nu} =  R^\sigma_{\,\,\,\mu}  R^\rho_{\,\,\,\nu} g_{\sigma\rho}
\end{array} \right .
\eeq{e:biconds}

\subsection{Supersymmetry variations}

In general one does not expect the boundary conditions derived as field
equations to be supersymmetric, since the action is
invariant under the supersymmetry algebra only up to a boundary term.
To ensure worldsheet supersymmetry the boundary conditions should make
the supersymmetry variation of the action vanish.  The constraints on
$R^\mu_{\,\,\nu}$ implied by this condition must then be made
compatible with (\ref{genfullNbc1}) and (\ref{RRcond}).

Assuming the standard (1,1) supersymmetry variation (\ref{compsusytr})
with supersymmetry parameter $\epsilon^+=\eta \epsilon^-$, the
variation of (\ref{acrbas}) yields the boundary term
\ber
\nonumber \delta_s S &=& \epsilon^{-} \int d\tau
\left [g_{\mu\nu} (\partial_\+ X^\mu \psi_-^\nu  -\eta
\partial_= X^\nu  \psi_+^\mu  ) +
 ig_{\mu\nu} F_{+-}^\mu(\psi_+^\nu +\eta \psi_-^\nu) - \right. \\
&& \left. - i \psi_+^\mu \psi_-^\nu \psi_+^\rho
 \Gamma_{\mu\nu\rho} -i \eta \psi_+^\mu\psi_-^\nu
 \psi_-^\rho \Gamma_{\nu\mu\rho}
 \right ] .
\eer{susyvar}
Inserting the ansatz (\ref{fermbc}), $\delta_s S$ simplifies to
\beq
\delta_s S= \eta \epsilon^{-} \int d\tau \psi_{+}^\mu \left [
  g_{\mu\nu} (R^\nu_{\,\,\sigma} \partial_{\+} X^\sigma - \partial_{=}
  X^\nu) + 2i \eta g_{\mu\nu} P^\nu_{\,\,\sigma} F_{+-}^\sigma - 2i
  \psi_{+}^\sigma \psi_{+}^\gamma R^\rho_{\,\,\sigma} P^\nu_{\,\,\mu}
  \Gamma_{\nu\rho\gamma} \right ].
\eeq{susyvar1} 

We now have two choices; either to stay completely off-shell, or to make
use of the (algebraic) bulk field equations for the auxiliary field
$F$.\footnote{Note that there are no boundary contributions in
  deriving the $F$-field equations.} If we stay off-shell and plug the
conditions (\ref{genfullNbc1}) into (\ref{susyvar1}) we get a
condition involving the $F$-field. However, this condition is not
uniquely determined; for example, one can always add a term of the
form $K_{\mu\nu\rho}\psi_+^\nu\psi_+^\rho$, where $K_{\mu\nu\rho}$ is
symmetric either in the first two indices or in the first and the last
(for instance, one could choose
$K_{\mu\nu\rho}=g_{\mu\sigma}\nabla_\nu R^\sigma_{\,\,\rho}$).
Therefore, staying off-shell, we do not obtain a unique form of the
boundary condition for $F$. (Even if this $F$-condition were unique,
we still need it to be compatible with the $F$-field equation on the
boundary.) Hence the way to proceed is to go partly on-shell and use the
$F$-field equation of motion,
\beq
g_{\mu\nu} F_{+-}^\nu + \psi_{+}^\nu \psi_{-}^\rho \Gamma_{\mu\nu\rho}
= 0,
\eeq{Ffieldeq}
restricted to the boundary (i.e., with the ansatz (\ref{fermbc}) inserted).
Substituting this in
(\ref{susyvar1}) yields the supersymmetry variation
$$
\delta_s S= \eta \epsilon^{-} \int d\tau \psi_{+}^\mu \left [
  g_{\mu\nu} (R^\nu_{\,\,\sigma} \partial_{\+} X^\sigma - \partial_{=}
  X^\nu) \right ],
$$
which, using the bosonic conditions from (\ref{genfullNbc1}),
reduces to
\beq
\delta_s S= i \eta \epsilon^{-} \int d\tau \psi_{+}^\mu \psi_+^\sigma
\psi_+^\gamma P^\delta_{\,\,\mu} R^\nu_{\,\,\sigma} g_{\nu\rho}
\nabla_\delta R^\rho_{\,\,\gamma}.
\eeq{susyvar3}
$\delta_s S$ vanishes only if $R^\mu_{\,\,\nu}$ satisfies
\beq
P^\delta_{\,\,\mu} R^\nu_{\,\,\sigma} g_{\nu\rho} \nabla_\delta
R^\rho_{\,\,\gamma} + P^\delta_{\,\,\sigma} R^\nu_{\,\,\gamma}
g_{\nu\rho} \nabla_\delta R^\rho_{\,\,\mu}+ P^\delta_{\,\,\gamma}
R^\nu_{\,\,\mu} g_{\nu\rho} \nabla_\delta R^\rho_{\,\,\sigma}=0,
\eeq{Rweakcond}
which, by contraction with $Q$, leads to
\beq
P^\sigma_{\,\,\gamma} P^\mu_{\,\,\nu} \nabla_{[\sigma}
Q^\delta_{\,\,\mu]} = P^\mu_{\,\,\gamma}
P^\nu_{\,\,\sigma} Q^\rho_{\,\,[\mu,\nu]} =0.
\eeq{itegrPform}
This is the \emph{integrability condition} for $P$ (cf.~Appendix~\ref{a:APM}).

In conclusion, the boundary conditions (\ref{genfullNbc1}) are
consistent with worldsheet supersymmetry, in the sense discussed
above, when $P$ is integrable.  The geometrical meaning of this
integrability condition is discussed in Section~\ref{s:geometric}.

\subsection{Compatibility with the algebra}
\label{s:algebra}

We may rephrase the question about boundary supersymmetry and study
compatibility with the supersymmetry \emph{algebra}.  This means that
the boundary conditions should be consistent with the supersymmetry
transformation of our fermionic ansatz (\ref{fermbc}). Applying the
transformations (\ref{compsusytr}) with $\epsilon^+=\eta\epsilon^-$
to the ansatz one finds a bosonic
boundary condition\footnote{The set of supersymmetric boundary
  conditions can be written in terms of 1D superfields, see
  Appendix~\ref{a:1D}.}
\beq
\d_{=} X^\mu =R^\mu_{\,\,\nu}\d_{\+}X^\nu + 2i\eta P^\mu_{\,\,\nu}
F_{+-}^\nu - 2iR^\mu_{\,\,\nu,\sigma} P^\sigma_{\,\,\rho}
\psi^\rho_{+} \psi^\nu_{+} .
\eeq{bbcNt}
Compatibility with the supersymmetry algebra is attained when the
boundary field equations (\ref{genfullNbc1})
satisfy (\ref{bbcNt}). Note that (\ref{bbcNt}) is stronger than the
requirement of zero supersymmetry variation of the action; this
may be seen by substituting (\ref{bbcNt}) into (\ref{susyvar1}) to get
\beq
\delta_s S= -2i \eta \epsilon^{-} \int d\tau\,\, \psi_{+}^\mu
\psi_{+}^\sigma \psi_{+}^\gamma g_{\sigma\nu} P^\rho_{\,\,\mu}
\nabla_\rho R^\nu_{\,\,\gamma},
\eeq{susyvar4}
which vanishes identically because $g_{\sigma\nu} P^\rho_{\,\,\mu}
\nabla_\rho R^\nu_{\,\,\gamma}$ is symmetric in $\sigma$ and $\gamma$
(by (\ref{RRcond})).

To show that the boundary equations of motion are consistent with
(\ref{bbcNt}), we first
go on-shell by using the $F$-field equation (\ref{Ffieldeq})
restricted to the boundary, obtaining
\beq
\d_= X^\mu - R^\mu_{\,\,\nu}\d_\+ X^\nu + 2i P^\sigma_{\,\,\gamma}
\nabla_\sigma R^\mu_{\,\,\nu} \psi_+^\gamma \psi_+^\nu = 0.
\eeq{algebFeqcon}
Contraction with $Q$ then gives the integrability condition for $P$,
after using that $Q^\mu_{\,\,\nu}\d_0X^\nu=0$.  On the other hand,
contracting (\ref{algebFeqcon}) with $P$ and using that $P$ is
integrable, we arrive at the second (bosonic) condition in
(\ref{genfullNbc1}).  Thus we conclude that (\ref{genfullNbc1})
\emph{are the general parity invariant boundary conditions compatible
  with the supersymmetry algebra}, provided that $R$ squares to one
and preserves the metric, and that $P$ is integrable.

\subsection{Preservation of the metric}
\label{s:gpreserve}

The properties (\ref{e:biconds}) allow us to draw some conclusions
about the form of $R^\mu_{\,\,\,\nu}$.  For a general metric
$g_{\mu\nu}$ there is only one solution for $R$ that squares to one
and preserves the metric, namely
$R^\mu_{\,\,\,\nu}=\delta^\mu_{\,\,\,\nu}$.  Thus there can be no
Dirichlet directions, i.e., this general background cannot support
D-branes, or worldsheet supersymmetry is broken.

If, on the other hand, the metric is not of a completely general form,
there may be other solutions for $R^\mu_{\,\,\,\nu}$, depending on the
form of $g_{\mu\nu}$. In the presence of D$p$-branes, the metric must
not mix Neumann and Dirichlet directions.  To see why, go to
adapted coordinates ($X^m,X^i$) at some point in the target space;
then $R=\diag(1,-1)$ at this point, and preservation of the metric,
$$
g_{\mu\nu} =  R^\sigma_{\,\,\,\mu}  R^\rho_{\,\,\,\nu} g_{\sigma\rho},
$$
tells us that the only backgrounds that allow
D$p$-branes are those satisfying \cite{Ooguri:1996ck}
\beq
 g_{im} =0.
\eeq{restrb}
Thus we see that, at the given point, the metric must be such that the
Neumann and Dirichlet directions decouple, if the strings attached to
the brane are to remain supersymmetric.

There remains the important question of whether the adapted system of
coordinates at a point can be extended to a neighbourhood along the
D$p$-brane. This is where integrability comes in; we address this issue
in Section~\ref{s:geometric}.

\subsection{A special case}
\label{s:case}

One may ask what is required for the two-fermion term in the
bosonic boundary conditions in (\ref{genfullNbc1}) to vanish, so that the
boundary conditions take the simple form
\beq
 \left \{ \begin{array}{l}
\psi_{-}^\mu = \eta R^\mu_{\,\,\nu} \psi_{+}^\nu \\
\d_{=} X^\mu =R^\mu_{\,\,\nu}\d_{\+}X^\nu
\end{array} \right .
\eeq{fullbcfaction} 
often assumed in the literature; see, e.g., \cite{Ooguri:1996ck}.
>From (\ref{genfullNbc1}) it is easily seen that
in addition to (\ref{e:biconds}) one needs to impose
\beq
P^\rho_{\,\,\mu} \nabla_\rho R^\nu_{\,\,\gamma} = 0, 
\eeq{RRprop}
a condition that also implies $P$-integrability
(c.f.~Eqn.~(\ref{itegrPform})).

Note that for a general metric, where we have
$R^\mu_{\,\,\,\nu}=\delta^\mu_{\,\,\,\nu}$, the conditions
(\ref{fullbcfaction}) reduce to
\beq
 \left \{ \begin{array}{l}
 \psi_{-}^\mu - \eta \psi_{+}^\mu = 0 \\
 \partial_{=}X^\mu - \partial_{\+} X^\mu =0 
\end{array} \right .
\eeq{fullNbc}
which corresponds to the freely moving open string,
its ends satisfying Neumann conditions in all directions
(corresponding to space-filling D9-branes).

\section{${\bf N=1}$ superconformal symmetry}
\label{s:currents}

The second route to finding the boundary conditions allowed by the
supersymmetric sigma model involves studying the conserved currents.
Here we derive these currents and the conditions they must satisfy on
the boundary, showing how this yields constraints on
$R^\mu_{\,\,\,\nu}$.

\subsection{Conserved currents}

We want to retain classical superconformal invariance in the presence
of boundaries. To do this, the appropriate objects to study are the
currents corresponding to supertranslations in $(1,1)$ superspace.
These supercurrents may be derived using superspace
notation (we sketch the main steps in Appendix~\ref{a:supercurrents}),
and we obtain
\beq
 T_{\+}^- = D_{+}\Phi^\mu \partial_{\+} \Phi^\nu g_{\mu\nu} ,
\eeq{superc1}
\beq
 T_{=}^+ = D_{-}\Phi^\mu \partial_{=} \Phi^\nu g_{\mu\nu} ,
\eeq{superc2}
obeying the conservation laws
$$
 D_{+}T_{=}^+ =0,\,\,\,\,\,\,\,\,\,\,\,\,\,\,\, D_{-}T_{\+}^- =0.
$$
Note that these conserved currents are defined only up to
the equations of motion.

The components of the supercurrents (\ref{superc1}) and
(\ref{superc2}) correspond to the supersymmetry
current and stress tensor as follows,
\beq
G_{+} =  T_{\+}^-| = \psi_{+}^\mu \d_{\+} X^\nu g_{\mu\nu} ,
\eeq{comp1}
\beq
 G_{-} = T_{=}^+|= \psi_{-}^\mu \d_{=} X^\nu g_{\mu\nu} ,
\eeq{comp2}
\beq
T_{++} = -i D_{+} T_{\+}^- |=\d_{\+}X^\mu \d_{\+}X^\nu g_{\mu\nu} +
 i \psi^\mu_+ \nabla_{+} \psi^\nu_{+} g_{\mu\nu} ,
\eeq{comp4}
\beq
T_{--} = - i D_{-} T_{=}^+ | = \d_{=}X^\mu \d_{=}X^\nu g_{\mu\nu} +
 i \psi^\mu_- \nabla_{-} \psi^\nu_{-} g_{\mu\nu} .
\eeq {TTdefinition}
The covariant derivative acting on the worldsheet fermions is given by
\beq
\nabla_{\pm}\psi_{+}^\nu = \partial_{\pp}\psi_{+}^\nu +
\Gamma^{\nu}_{\,\,\rho\sigma}\d_{\pp} X^\rho
\psi_{+}^\sigma,\,\,\,\,\,\,\,\,\,\, \nabla_{\pm}\psi_{-}^\nu =
\partial_{\pp}\psi_{-}^\nu + \Gamma^{\nu}_{\,\,\rho\sigma}\d_{\pp}
X^\rho \psi_{-}^\sigma ,
\eeq{covdervferm}
where $\Gamma^{\nu}_{\,\,\rho\sigma}$ is the Levi-Civita symbol.
Moreover, the conservation laws in components
acquire the following form,
$$
\begin{array}{r@{\hspace{2cm}}r}
\d_\+ G_- =0 & \d_\+ T_{--}=0 \\
\d_= G_+ = 0 & \d_= T_{++} = 0 
\end{array}
$$

To ensure superconformal symmetry on the boundary
we need to impose boundary conditions on the
currents (\ref{comp1})--(\ref{TTdefinition}). To see what
these conditions look like, consider the conserved charge,
\beq
0=\partial_0 Q = \int d\sigma \,\,\, \partial_0 J^0,
\eeq{e:charge}
where the current $J$ is any of the currents $G$ and
$T$, and $J^0$ is the $\tau$-component of $J$.
Then current conservation, $\partial_\alpha J^\alpha =0$,
implies that
$$
0= - \int d\sigma \,\,\, \partial_1 J^1,
$$
resulting in the boundary condition
$$
 J^{+}-J^{-} =0.
$$
Applying this to our currents $G$ and $T$, we arrive at the boundary
conditions\footnote{The $\eta$ appears here due to supersymmetry.}
\beq
 G_{+}-\eta G_{-} =0,\,\,\,\,\,\,\,\,\,\,\,\,\,\,\,\,\,
 T_{++}-T_{--}=0.
\eeq{superconf}
At the classical level, these conditions are just
saying that the left-moving super-Virasoro algebra coincides with the
right-moving one. We emphasize that classically
the conditions (\ref{superconf}) can make sense only on-shell.  This
means that we may (and should) make use of the field equations in our
analysis of the current constraints.

\subsection{Boundary conditions}

To examine how the current conditions affect the choice of
$R^\mu_{\,\,\nu}$ in the fermionic boundary conditions, we begin by
rewriting the stress tensor in a suitable form.  The problem is that
the stress tensor has both normal and tangential fermionic
derivatives. This is remedied by defining $2\nabla_0 \equiv \nabla_{+}
+ \nabla_{-}$ and using the fermionic equations of motion,
$$
g_{\mu\nu} (\psi_+^\mu \nabla_- \psi_+^\nu
- \psi_-^\mu \nabla_+ \psi_-^\nu) =0,
$$
to rewrite $T_{++}-T_{--}$ as follows,
\ber
\nonumber T_{++}-T_{--} &=& \d_\+ X^\mu \d_\+ X^\nu g_{\mu\nu} - \d_=
X^\mu \d_= X^\nu g_{\mu\nu} + \\
\nonumber && + i (\psi_{+}^\mu - \eta\psi_-^\mu) \nabla_0
(\psi_+^\nu + \eta \psi_-^\nu) g_{\mu\nu} + \\
&& + i (\psi_+^\mu + \eta \psi_-^\mu)\nabla_0
(\psi_+^\nu-\eta \psi_-^\nu) g_{\mu\nu} .
\eer{TTTnew}
The general form of the boundary
conditions which satisfy (\ref{superconf}) are found by again making
the ansatz (\ref{fermbc}),
\beq
 \psi_{-}^\mu = \eta R^\mu_{\,\,\nu} \psi_{+}^\nu,\,\,\,\,\,\,\,\,\,\,\,\,\,\,
 R^\mu_{\,\,\nu} R^\nu_{\,\,\sigma} = \delta^\mu_{\,\,\sigma},
\eeq{fbcondRR}
recalling that the last property is needed if we
want worldsheet parity as a symmetry of the boundary
conditions. Using (\ref{TTTnew}) and (\ref{fbcondRR}),
the conditions (\ref{superconf}) imply the following two properties,
\beq
 g_{\mu\nu} =  R^\sigma_{\,\,\,\mu}  R^\rho_{\,\,\,\nu} g_{\sigma\rho},
\,\,\,\,\,\,\,\,\,\,\,\,\,\,
P^\sigma_{\,\,\gamma} P^\mu_{\,\,\nu} \nabla_{[\sigma}
Q^\delta_{\,\,\mu]} = 0,
\eeq{RRcond2}
i.e., $R$ preserves the metric and $P$ is integrable, just
as we saw in Section~\ref{s:action}. The first of these
properties is straightforwardly obtained from the
$T$-condition; the second is obtained by using the first property to
rewrite the $G$- and $T$-conditions as
\beq
\left \{ \begin{array}{ll}
    G_{+}-\eta G_{-} =& \psi_+^\sigma (
    g_{\sigma\nu} \d_\+ X^\nu
    - R^\mu_{\,\,\sigma} \d_= X^\nu g_{\mu\nu}) =0 \\
    T_{++}-T_{--} =& 2\d_0 X^\rho \left (g_{\rho\nu} (\d_\+ X^\nu -
      \d_= X^\nu) - i P^\delta_{\,\,\rho} R^\mu_{\,\,\sigma}
      g_{\mu\nu} \nabla_\delta R^\nu_{\,\,\gamma} \psi_{+}^\sigma
      \psi_+^\gamma \right ) =0
\end{array} \right .
\eeq{GGTT1}
and inserting $\delta^\mu_{\,\,\nu} = P^\mu_{\,\,\nu}+
Q^\mu_{\,\,\nu}$ between $\d_0 X^\rho$ and the parenthesis in the
$T$-condition (we assume that there are Dirichlet directions, i.e.,
that there is a non-vanishing $Q$ such that $Q^\mu_{\,\,\nu}\d_0 X^\nu
=0$). We then use the result in the $G$-condition, thus obtaining
the second of Eqns~(\ref{RRcond2}).

A remark may be in order on the fact that we obtain the individual
conditions (\ref{RRcond2}), rather than some more general condition
involving all the worldsheet fields. This may be understood by
recognising that, after using the fermionic boundary condition and the
current conditions, we have reduced the original set of four
independent fields $\left\{ \psi_{\pm}^\mu, \d_{\pp} X^\mu \right\}$
to a set of only two independent fields, say $\psi_{+}^\mu$ and $\d_\+
X^\mu$. But since these remaining fields are truly independent, terms
of different structure that appear in the current conditions must
vanish separately. Thus, for example in the analysis of the
$G$-condition, a term
$$
\psi_{+}^\mu \d_\+ X^\nu (g_{\mu\nu} -
R^\sigma_{\,\,\,\mu} R^\rho_{\,\,\,\nu} g_{\sigma\rho})
$$
will appear which must vanish independently, giving the first of
Eqns~(\ref{RRcond2}), and the second condition is obtained
analogously.

We conclude that the complete set of boundary conditions
that solve the current boundary conditions (\ref{superconf}) are
\beq
\left \{ \begin{array}{l}
    \psi_{-}^\mu - \eta R^\mu_{\,\,\nu} \psi_{+}^\nu = 0 \\
    g_{\delta\mu} P^\mu_{\,\,\nu}( \partial_{\+}X^\nu - \partial_{=}
    X^\nu) - i P^\mu_{\,\,\delta} R^\nu_{\,\,\sigma}g_{\nu\rho}
    \nabla_\mu R^\rho_{\,\,\gamma}
    \psi_{+}^\sigma \psi_+^\gamma =0 \\
P^\sigma_{\,\,\gamma} P^\mu_{\,\,\nu} \nabla_{[\sigma}
Q^\delta_{\,\,\mu]} = 0 \\
    Q^\mu_{\,\,\nu}(\partial_{=}X^\nu + \partial_{\+} X^\nu) =0 \\
 g_{\mu\nu} =  R^\sigma_{\,\,\,\mu}  R^\rho_{\,\,\,\nu} g_{\sigma\rho} \\
R^\mu_{\,\,\,\nu}  R^\nu_{\,\,\,\rho} = \delta^\mu_{\,\,\,\rho}
\end{array} \right .
\eeq{newTTT}
These conditions are identical to the boundary conditions
(\ref{genfullNbc1}), (\ref{e:biconds}) derived from the action in
Section~\ref{s:action}.  In particular, it is clear that
again $P$ needs to be integrable.

As an alternative to the above procedure, one may derive the
conditions (\ref{newTTT}) by imposing compatibility with the
supersymmetry algebra on the currents in a more direct manner.  That
is, we can use the (on-shell) supersymmetry transformation
(\ref{algebFeqcon}) of the fermionic boundary conditions. The result
is again that the $G$- and $T$-conditions require, in addition to the
fermionic boundary condition and its supersymmetry transformation,
that $R^\mu_{\,\,\,\nu}$ satisfy (\ref{RRcond2}), i.e., we arrive at
the boundary conditions (\ref{newTTT}).  Note that now
$P$-integrability may also be derived in the same way as discussed in
Section~\ref{s:algebra}, contracting (\ref{algebFeqcon}) with $Q$.

\section{Geometric interpretation}
\label{s:geometric}

In this section we discuss the target space interpretation of the
boundary conditions we have found.  All information about the boundary
conditions is encoded in $R^\mu_{\,\,\nu}$, so we focus on the
restrictions on $R^\mu_{\,\,\nu}$ alone.  There are two ways of
viewing the boundary conditions, locally and globally; either
$R^\mu_{\,\,\nu}$ is defined only locally in the target space, or it
is defined globally.

\subsection{Locally defined conditions}

We first take the local point of view, assuming that $R^\mu_{\,\,\nu}$
is defined only in some region of the target space manifold.  The
first boundary condition we consider is the integrability of $P$.
Take an arbitrary point $x^\mu_0$ in the region where
$R^\mu_{\,\,\nu}$ is defined.  We may then write any contravariant
vector $dX^\mu$ at $x^\mu_0$ as
\beq
 dX^\mu = P^\mu_{\,\,\nu}dX^\nu + Q^\mu_{\,\,\nu}dX^\nu,
\eeq{dXexpan}
where we have used the fact that $\delta^\mu_{\,\,\nu} =
P^\mu_{\,\,\nu} + Q^\mu_{\,\,\nu}$. It follows that
the distribution $P$ is defined by
\beq
 Q^\mu_{\,\,\nu}(X)dX^\nu =0,
\eeq{definP}
since this leaves
\beq
 dX^\mu = P^\mu_{\,\,\nu}dX^\nu.
\eeq{dXPdist}
The definition (\ref{dXPdist}) leads to $P$-integrability, by acting
on it with the exterior derivative $d$,
\beq
 0=d^2X^\mu = P^\mu_{\,\,\nu,\rho} dX^\rho \wedge dX^\nu
= \frac{1}{2} P^\mu_{\,\,[\nu,\rho]} P^\rho_{\,\,\lambda} P^\nu_{\,\,\sigma}
 dX^\lambda \wedge dX^\sigma,
\eeq{ddXP}
whence
$$
P^\rho_{\,\,\lambda} P^\nu_{\,\,\sigma} Q^\mu_{\,\,[\nu,\rho]} =0,
$$
which is the $P$-integrability condition
(cf.~Eqn.~(\ref{integrablP})).

It turns out that $P$-integrability is the very condition necessary
and sufficient for the differential equations (\ref{definP}) to be
completely solvable in a neighbourhood. To see this, note that for
$Q^\mu_{\,\,\nu}(X)dX^\nu=0$ to be integrable, it is necessary
and sufficient that in a neighbourhood,
\beq
0 = d(Q^\mu_{\,\,\nu}(X)dX^\nu) = \frac{1}{2} Q^\mu_{\,\,[\nu,\rho]}
dX^\rho \wedge dX^\nu =\frac{1}{2} P^\rho_{\,\,\lambda} P^\nu_{\,\,\sigma}
Q^\mu_{\,\,[\nu,\rho]}dX^\lambda \wedge dX^\sigma .
\eeq{inePcondDre}

If the equations (\ref{definP}) are completely integrable, then they
admit $\rank(Q)$ independent solutions.
Going to a coordinate basis where the equations take the form
$d\tilde{X}^i=0$, we may write the solutions as
\beq 
\tilde{X}^i(X)=q^i, \quad i=0,...,\rank(Q)-1,
\eeq{solQequat}
where $q^i$ are constants.

Interpreting the above result in terms of the target space manifold,
we see that the coordinates (\ref{solQequat}) define a
$\rank(P)$-dimensional submanifold. If $P$ has rank $p+1$, then the
submanifold is ($p+1$)-dimensional, i.e., it is a D$p$-brane, and
$\{\tilde{X}^i\}$ is the system of coordinates adapted to the brane.
Now $P$-integrability allows us to extend this system of coordinates
to a neighbourhood (i.e., it is defined not only at one point) along
the Neumann directions.  Note that this result arises in a purely
algebraic fashion from the requirement of minimal worldsheet
supersymmetry (see Section~\ref{s:action}).

Next we interpret the requirement that $R$ preserve the metric,
\beq
g_{\mu\nu} =
R^\sigma_{\,\,\,\mu} R^\rho_{\,\,\,\nu} g_{\sigma\rho}.
\eeq{e:gRR}
Unlike $P$-integrability, this property is not purely algebraic; it
requires additional information, for instance conserved currents
(see Section~\ref{s:currents}).

As we saw in Section~\ref{s:gpreserve}, preservation of the metric
implies that in coordinates adapted to the brane the metric must take
a block diagonal form, $g_{\mu\nu} = \diag(g_{mn},g_{ij})$, on the
worldsheet boundary.  Given that $P$ is integrable, we can extend the
adapted coordinate system to a neighbourhood along the brane, so that
$R=\diag(1,-1)$ in this neighbourhood. Thus the metric is block
diagonal in this neighbourhood, and the metric along the Neumann
directions, $g_{nm}$, could in principle serve as a metric on the
D$p$-brane.

In conclusion, we see that requiring $N=1$ superconformal invariance
for open strings results in that they are allowed to end only
on Riemannian submanifolds of the target space manifold.  Of course,
this implies no restrictions on the ``bulk'' target space; the
conditions here only applies to the worldsheet boundary, telling us
where the strings may end, not what the rest of the background looks
like. As an aside, note that if applied to the case where both $P$ and
$Q$ are integrable, the above discussion leads to the existence of
both D$p$- and D$(d-(p+1)-1)$-branes, in a $d$-dimensional target space.

\subsubsection{Confined geodesics}

As we discussed in Section~\ref{s:case}, the fermionic boundary
condition and its supersymmetry transformation simplify to
(\ref{fullbcfaction}) when the extra requirement (\ref{RRprop}) is
imposed. This requirement has a nice geometrical interpretation
when defined in a neighbourhood of a brane. Since
$\nabla_\rho R^\nu_{\,\,\gamma} = - 2 \nabla_\rho
Q^\nu_{\,\,\gamma}$, one can rewrite (\ref{RRprop}) as
\beq
  P^\rho_{\,\,\mu} \nabla_\rho Q^\nu_{\,\,\gamma} = 0 . 
\eeq{Pparl}
When this condition is satisfied, $P$ is called \emph{parallel} \cite{Yano1}
with respect to the Levi-Civita connection. Indeed, if $P$ is
integrable one can always construct a symmetric affine connection such
that (\ref{Pparl}) holds.

To understand the physical meaning of (\ref{Pparl}), take a point
$x^\mu_0$ and a vector $v^\mu$ at $x^\mu_0$, which is contained in $P$
(i.e., $Q^\mu_{\,\,\nu}v^\nu=0$). Then the autoparallel curve with
respect to the Levi-Civita connection, $v^\mu\nabla_\mu v^\nu=0$, is
uniquely determined by the initial point $x^\mu_0$ and the initial
direction $v^\mu$. The condition (\ref{Pparl}) ensures that the path
thus determined is always contained in $P$. This is easily seen by
inserting  $v^\mu =P^\mu_{\,\,\nu}v^\nu$ into the curve (where
now $v^\mu$ is any vector along the curve),
\begin{eqnarray*}
0=v^\mu\nabla_\mu v^\nu &=& v^\mu \nabla_\mu
\left( P^\nu_{\,\,\sigma} v^\sigma \right) \\
&=& v^\mu v^\sigma \nabla_\mu P^\nu_{\,\,\sigma} +
P^\nu_{\,\,\sigma} v^\mu \nabla_\mu v^\sigma \\
&=& - v^\lambda v^\sigma P^\mu_{\,\,\lambda} \nabla_\mu
Q^\nu_{\,\,\sigma} .
\end{eqnarray*}
Thus the autoparallel curve is compatible
with the requirement that $v^\mu$ be contained in $P$,
if (\ref{Pparl}) holds.

A distribution $P$ such that autoparallel curves starting out in $P$
always remain in $P$ is called \emph{geodesic}\footnote{This property
  is a weaker condition than (\ref{Pparl}).} \cite{Yano1}.  Physically
it means that if a geodesic starts on a D$p$-brane, then it will
always remain on the brane. Hence particles cannot escape from the
brane.

\subsection{Globally defined conditions}

Turning now to global issues, we ask when it makes sense to interpret
the boundary conditions (\ref{genfullNbc1}), (\ref{e:biconds})
globally and what kind of restrictions it implies for the target
space.  If the tensor $R^\mu_{\,\,\nu}$ is defined globally on the
target space, and if the property $R^\mu_{\,\,\nu}R^\nu_{\,\,\rho}=
\delta^\mu_{\,\,\rho}$ holds globally, then the target space manifold
is said to admit an \emph{almost product structure} $R^\mu_{\,\,\nu}$ (see
Appendix~\ref{a:APM}). If in addition $R^\mu_{\,\,\nu}$ preserves the
metric, then the manifold is called an \emph{almost product Riemannian
manifold}.

Thus our boundary conditions tell us that when the target space is an
almost product Riemannian manifold with the property that $P$ is
integrable, then it admits D$p$-brane solutions at any point, and that
these solutions respect $N=1$ superconformal symmetry.  If the target
space is a \emph{locally product manifold}, i.e., both $P$ and $Q$ are
integrable, then the sigma model admits both D$p$- and
D$(d-(p+1)-1)$-brane solutions at any point of the manifold, and they
are all compatible with $N=1$ superconformal invariance.

\subsubsection{Warped product spacetimes}

Consider again the case where $P$ is a geodesic distribution
satisfying (\ref{Pparl}). There is an interesting example of a geometry
for which this condition is fulfilled globally, namely the warped
product spacetimes \cite{Yano1}. Given two Riemannian manifolds
$({\cal M}_i, g_i)$, $i=1,2$, and a smooth function $f:{\cal M}= {\cal
  M}_1 \times {\cal M}_2 \rightarrow \bR$, construct the metric $g=g_1
\oplus e^{f} g_2$ on ${\cal M}$. Then the manifold (${\cal M}, g$) is
a \emph{warped product manifold}.\footnote{In the present context it
  is interesting that many exact solutions of Einstein equations,
  e.g., Schwarzschild, Robertson-Walker, Reissner-Nordstr\"om, etc.,
  and also $p$-brane solutions, are examples of warped product
  spacetimes.} Thus, if (\ref{Pparl}) is defined globally, we know
that the target space is a warped product spacetime, and we also know
from Section~\ref{s:case} that the only boundary conditions preserving
$N=1$ superconformal symmetry in this case are
$$
\psi^\mu_- = \eta R^\mu_{\,\,\nu} \psi^\nu_+ ,
\,\,\,\,\,\,\,\,\,\,\,
\d_= X^\mu = R^\mu_{\,\,\nu} \d_\+ X^\nu ,
$$
although now they need to be globally defined.

\section{Additional supersymmetry}
\label{s:moresusy}

In this section we discuss the non-linear sigma model with $N=2$
supersymmetry.  Although our discussion is somewhat sketchy we show
that the solution of the problem is more general than usually
assumed in the literature.

We have shown that for the $N=1$ supersymmetric sigma model (\ref{acrbas})  
 the following set of parity invariant boundary conditions,
\beq
\left \{ \begin{array}{l}
  \psi_-^\mu - \eta R^\mu_{\,\,\nu} \psi^\nu_+ =0 , \\
  \d_= X^\mu - R^\mu_{\,\,\nu} \d_\+ X^\nu + 2i P^\sigma_{\,\,\gamma} \nabla_\sigma
 R^\mu_{\,\,\nu}\psi_+^\gamma \psi_+^\nu =0 ,
\end{array} \right .
\eeq{genbc2}  
respect $N=1$ superconformal invariance. The fermionic and bosonic
boundary conditions are related to each other through supersymmetry
transformations.  Due to the integrability of $P$ this solution admits
a nice geometrical interpretation in terms of a submanifold.

Now if the model (\ref{acrbas}) has a second supersymmetry the target
space must
be a K\"ahler manifold. In real coordinates the second on-shell supersymmetry 
 can be written as follows,
\beq
\left \{\begin{array}{l}
 \delta_2 X^\mu = - (\epsilon^+_2 \psi_+^\nu + \epsilon_2^- \psi_-^\nu) J^\mu_{\,\,\nu} \\
 \delta_2 \psi_+^\mu = -i \epsilon_2^+ \d_\+ X^\nu J^\mu_{\,\,\nu} + \epsilon_2^- 
 J^\mu_{\,\,\sigma}\Gamma^{\sigma}_{\,\,\nu\rho} \psi_-^\rho \psi_+^\nu 
 + \epsilon_2^+ J^\mu_{\,\,\nu,\rho} \psi_+^\nu \psi_+^\rho + \epsilon_2^- J^\mu_{\,\,\nu,\rho}
 \psi_+^\nu \psi_-^\rho  \\
 \delta_2 \psi_-^\mu = -i\epsilon_2^- \d_= X^\nu J^\mu_{\,\,\nu} 
 - \epsilon_2^+ J^\mu_{\,\,\sigma}\Gamma^{\sigma}_{\,\,\nu\rho}
 \psi_-^\rho \psi_+^\nu +\epsilon_2^+ J^\mu_{\,\,\nu,\rho} \psi_-^\nu \psi_+^\rho
 + \epsilon_2^- J^\mu_{\,\,\nu,\rho}\psi_-^\nu \psi_-^\rho
\end{array}
\right .
\eeq{n11}
where $J^\mu_{\,\,\nu}$ is a complex structure (i.e.,
$J^\mu_{\,\,\nu} J^\nu_{\,\,\rho} = - \delta^\mu_{\,\,\rho}$ and $J$
is integrable) with the property $\nabla_\rho
J^\mu_{\,\,\nu}=0$. The new symmetry
gives rise to several new questions. For instance,
what are the boundary conditions that ensure
survival of this symmetry on the boundary? And what restrictions, if
any, does it imply for the conditions (\ref{genbc2})? Here we
address the latter of these questions. To this end, we start from the
fermionic boundary condition in (\ref{genbc2}) and use the second
supersymmetry variation to derive the corresponding bosonic boundary
conditions,
\beq
 \d_= X^\mu + (\eta\eta_2)J^\mu_{\,\,\sigma} R^\sigma_{\,\,\nu} J^\nu_{\,\,\gamma} 
 \d_\+ X^\gamma - i\left [ (\eta\eta_2) J^\mu_{\,\,\sigma}
 \nabla_\rho R^\sigma_{\,\,\nu} J^\rho_{\,\,\gamma} +
   J^\mu_{\,\,\sigma} \nabla_\rho R^\sigma_{\,\,\nu} J^\rho_{\,\,\lambda} 
 R^\lambda_{\,\,\gamma}  \right ] \psi_+^\gamma \psi_+^\nu =0,
\eeq{susy2aa}
where we assume that $\epsilon_2^+=\eta_2\epsilon_2^-$.  The equations
(\ref{genbc2}) and (\ref{susy2aa}) should be equivalent.  Comparing
the $X$-part we get the following condition (which also is part of the
requirement for the second boundary supersymmetry),
\beq
 J^\mu_{\,\,\gamma} R^\gamma_{\,\,\nu}
 J^\nu_{\,\,\sigma} = - (\eta\eta_2) R^\mu_{\,\,\sigma} ,
\eeq{compbos} 
 or equivalently,
\beq
 J^\mu_{\,\,\gamma} R^\gamma_{\,\,\nu} = (\eta\eta_2) R^\mu_{\,\,\gamma} J^\gamma_{\,\,\nu}.
\eeq{compbos2}
The case $\eta\eta_2=1$ corresponds to the so-called B-type conditions
and $\eta\eta_2=-1$ to A-type conditions as defined in
\cite{Ooguri:1996ck}.  One can show that the B-type conditions
correspond to K\"ahler submanifolds and the A-type conditions to
 coisotropic submanifolds. We emphasize that integrability of $P$ plays
a crucial role in this argument.

Using the property (\ref{compbos}) we rewrite
 the equation (\ref{susy2aa}) as follows,
\beq
 \d_= X^\mu - R^\mu_{\,\,\nu}  
 \d_\+ X^\nu - i (\eta\eta_2)  J^\mu_{\,\,\lambda} J^\rho_{\,\,\gamma} \left [
 \nabla_\rho R^\lambda_{\,\,\nu}  +
   \nabla_\sigma R^\lambda_{\,\,\nu}  
 R^\sigma_{\,\,\rho}  \right ] \psi_+^\gamma \psi_+^\nu =0 .
\eeq{susy2bb}
 Comparing the two-fermion terms in (\ref{genbc2}) and (\ref{susy2bb}) we get
 the following condition,
\beq
 \left ( P^\rho_{\,\,\gamma} \nabla_\rho R^\mu_{\,\,\nu} + (\eta\eta_2) 
 J^\mu_{\,\,\lambda} J^\rho_{\,\,\gamma} P^\sigma_{\,\,\rho}  \nabla_\sigma R^\lambda_{\,\,\nu}  
 \right ) \psi_+^\gamma \psi_+^\nu =0  .
\eeq{condtwo}
This condition does not imply that the two-fermion term
in the bosonic boundary condition vanishes altogether.
However, it does put some restrictions on this term.

As an example we consider the B-type conditions. In this case it is
convenient to go to complex coordinates such that
$J^i_{\,\,j}=i\delta^i_{\,\,j}$, $J^{\bar{i}}_{\,\,\bar{j}}= -
i\delta^{\bar{i}}_{\,\,\bar{j}}$ and $J^{i}_{\,\,\bar{j}} =
J^{\bar{i}}_{\,\,j} =0$.  In these coordinates the property
(\ref{compbos}) implies
\beq
 R^i_{\,\,\bar{j}} = R^{\bar{i}}_{\,\,j}=0,\,\,\,\,\,\,\,\,\,\,\,\,\,
 \nabla_\rho R^i_{\,\,\bar{j}} = \nabla_\rho R^{\bar{i}}_{\,\,j}=0.
\eeq{holR}
The conditions (\ref{condtwo}) take the form
\beq
 P^{\bar{s}}_{\,\,\bar{l}} \nabla_{\bar{s}} R^i_{\,\,j} \psi_+^{\bar{l}} \psi_+^j=0,
\,\,\,\,\,\,\,\,\,\,\,\,\,\,\,
P^{s}_{\,\,l} \nabla_{s} R^{\bar{i}}_{\,\,\bar{j}}
 \psi_+^{l} \psi_+^{\bar{j}}= 0 ,
\eeq{fermcond}
whence the bosonic boundary condition may be written as
\beq
 \d_= X^i - R^i_{\,\,j} \d_\+ X^j + 2i P^s_{\,\,l} \nabla_s
 R^i_{\,\,j}\psi_+^l \psi_+^j =0  ,
\eeq{holbc}
in holomorphic coordinates.
The structure in front of the
two-fermion term is related to the second fundamental
 form of a submanifold \cite{now}.

We would like to stress that the solutions we have discussed are more
general than those usually considered in the literature (see
e.g.~\cite{Ooguri:1996ck} and \cite{Hori:2000ck}) and that they admit
a nice geometrical interpretation.
As one could see from our previous discussion, the requirements of zero
variation (\ref{fullvar}) and of conformal invariance on the boundary
(i.e., $T_{++}-T_{--}=0$) are completely equivalent. However, in the
literature one encounters restrictions on the boundary that are
stronger than the vanishing of (\ref{fullvar}).  For instance, in
\cite{Hori:2000ck} the following conditions are imposed on the
boundary,
\beq
 \begin{array}{l}
\delta X^\mu \,\, g_{\mu\nu} \,\, \d_1X^\nu =0 , \\
 g_{\mu\nu} (\delta \psi_+^\mu \psi_+^\nu - \delta \psi_-^\mu
\psi_-^\nu)=0 .
\end{array}
\eeq{vafacond}
(Note that in
normal coordinates the two-fermion term in (\ref{fullvar}) is absent.)
This is certainly a solution of the problem, but not the most general one.
The topic deserves further study and we plan to return to it in future work.

\section{Conclusions}
\label{s:discussion}

In this paper we have investigated what boundary conditions on an $N=1$
sigma model are required for supersymmetry.

First, vanishing of the boundary supersymmetry variations together
with invariance of the action under general variations give us a set
of boundary conditions. These turn out to be compatible with the
supersymmetry algebra on the boundary in that the fermionic and
bosonic boundary conditions form a supersymmetry multiplet on the
(auxiliary) $F$-field shell.

We further studied the boundary conditions required for left and right
currents to agree on the boundary. The conditions derived in this way
are identical to those we find using our first method.

Our boundary conditions are more general than those usually adopted in
the literature, and they are derived here in a systematic manner. We
believe that this makes them useful.

One interesting feature of our results is the occurrence of a mixed
second rank tensor $R^\mu_{\,\,\,\nu}$ that squares to the identity
and preserves the metric.  We have shown that the projector in the Neumann
directions, formed from $R^\mu_{\,\,\,\nu}$, satisfies an
integrability condition. This condition has the natural interpretation
that it is possible to choose coordinates along a D-brane such that
$R^\mu_{\,\,\,\nu}$ is constant and diagonal.  Our boundary conditions
give information about how D-branes may be embedded in spacetime and
what the corresponding local geometry looks like.  In particular, we
showed that minimal boundary supersymmetry requires open strings to end
only on (pseudo) Riemannian submanifolds of the target space.  This
fact is usually assumed in the literature, but we have derived it in a
rigorous way.

Mathematically it may also be of interest to consider the case when
$R^\mu_{\,\,\,\nu}$ is globally defined. Then it has the geometric
interpretation of an almost product structure, and we briefly
discussed this.

In this paper we have only considered sigma models in a non-trivial
background metric. The boundary conditions will get modified if a
background antisymmetric $B$-field is also included
\cite{Haggi-Mani:2000uc}. We turn to this case in a future
publication \cite{now}.

Other questions of interest to us are related to treating the full
theory, i.e., to include the 2D supergravity fields. And, even in the
gauge-fixed case, to include contributions from the ghost fields,
e.g., to the currents.

\bigskip

\bigskip

{\bf Acknowledgements}:
We are grateful to Ingemar Bengtsson and Andrea
Cappelli for discussions and comments. 
 MZ would like to thank
 the ITP, Stockholm University, where part of this work was carried out.
UL acknowledges support in part
by EU contract HPNR-CT-2000-0122 and by NFR grant 650-1998368.

\appendix

\section{(1,1) supersymmetry}
\label{a:11susy}

Throughout the paper we use $\mu,\nu,...$ as spacetime indices, $(\+,
=)$ as worldsheet indices, and $(+,-)$ as two-dimensional spinor
indices.  We also use superspace conventions, where the pair of spinor
coordinates of two-dimensional superspace are labelled $\th^{\pm}$,
and the covariant derivatives $D_\pm$ and supersymmetry generators
$Q_\pm$ satisfy
\ber
D^2_+ &=&i\d_\+, \quad
D^2_- =i\d_= , \quad \{D_+,D_-\}=0 , \cr
Q_\pm &=& -D_\pm+2i\th^{\pm}\d_{\pp} ,
\eer{alg}
where $\d_{\pp}=\partial_0\pm\partial_1$.  In terms of the covariant
derivatives, a supersymmetry transformation of a superfield $\P$ is
then given by
\ber
\delta \P &\equiv & (\e^+Q_++\e^-Q_-)\P \cr
&=& -(\e^+D_++\e^-D_-)\P
+2i(\e^+\th^+\d_\++\e^-\th^-\d_=)\P .
\eer{tfs}
The components of a superfield $\P$ are defined via projections as
follows,
\ber
\P|\equiv X, \quad D_\pm\P| \equiv \p_\pm, \quad D_+D_-\P|\equiv F_{+-}
,
\eer{comp}
where a vertical bar denotes ``the $\th =0$ part of ''.
Thus, in components, the $(1,1)$ supersymmetry transformations are given by
\beq
\left \{ \begin{array}{l}
    \delta X^\mu = - \epsilon^{+} \psi_+^\mu - \epsilon^- \psi_-^\mu \\
    \delta \psi_+^\mu =  -i\epsilon^+ \d_{\+}X^\mu - \epsilon^- F^\mu_{-+}\\
    \delta \psi_-^\mu  = -i \epsilon^- \d_{=} X^\mu - \epsilon^+ F_{+-}^\mu \\
    \delta F^\mu_{+-} = - i \epsilon^+ \d_{\+} \psi_-^\mu + i
    \epsilon^- \d_- \psi_+^\mu
\end{array} \right .
\eeq{compsusytr}

\section{1D superfield formalism}
\label{a:1D}

One may view the 2D supersymmetry algebra as a combination of two 1D
algebras (for similar considerations see \cite{Hori}). To see this, we
rewrite the $(1,1)$ supersymmetry algebra in terms of 1D
supermultiplets.  Assuming that $\epsilon^+ = \eta \epsilon$ and
$\epsilon^-\equiv \epsilon$, (\ref{compsusytr}) becomes
\beq 
\left \{
\begin{array}{l}
  \delta X^\mu = - \epsilon (\eta \psi_+^\mu + \psi_-^\mu) \\
  \delta (\eta \psi_+^\mu + \psi_-^\mu) = 
 -2i\epsilon \partial_0 X^\mu \\
 \delta ( \psi_-^\mu - \eta \psi_+^\mu ) = -2\epsilon
 ( \eta F_{+-}^\mu - i\partial_1 X^\mu ) \\
 \delta (\eta F_{+-}^\mu -i\partial_1 X^\mu ) =
 -i \epsilon \partial_0 (  \psi_-^\mu - \eta \psi_+^\mu)
\end{array} \right.  
\eeq{susyrew1}
Introducing a new notation for the following
combinations of fields,
\beq
\Psi^\mu \equiv \frac{1}{\sqrt{2}}(\eta \psi_+^\mu +
\psi_-^\mu),\,\,\,\,\,\,\,\,\, \tilde{\Psi}^\mu \equiv
\frac{1}{\sqrt{2}}( \psi_-^\mu -\eta \psi_+^\mu ),\,\,\,\,\,\,\,\,\,
f^\mu \equiv \eta F_{+-}^\mu -i \partial_1 X^\mu,
\eeq{newnot}
and redefining $\epsilon= \sqrt{2}\epsilon$,
the algebra (\ref{susyrew1}) takes on the simple form
\beq 
\left \{
\begin{array}{l}
  \delta X^\mu = - \epsilon \Psi^\mu \\
  \delta \Psi^\mu =  - i\epsilon \partial_0 X^\mu \\
  \delta \tilde{\Psi}^\mu = -\epsilon  f^\mu \\
  \delta f^\mu =  - i \epsilon \partial_0 \tilde{\Psi}^\mu 
\end{array} \right.  
\eeq{susyrew2}
Clearly, (\ref{susyrew2}) is a decomposition of the
2D algebra into two 1D supermultiplets.
We introduce a 1D superfield notation for these multiplets,
\beq
 K^\mu = X^\mu +\theta \Psi^\mu,\,\,\,\,\,\,\,\,\,\,\,\,\,\,\,\,
 S^\mu = \tilde{\Psi}^\mu + \theta f^\mu,
\eeq{1dsusyf}
where $\theta$ is the single Grassmann coordinate of the respective 1D
superspace, and the corresponding 1D superderivative is now $D$,
satisfying $D^2=i\partial_0$.

The fermionic boundary condition (\ref{fermbc}) and its supersymmetry
transformation may be rewritten in terms of the
1D supermultiplets,
\beq
\left\{ \begin{array}{l}
    \Psi^\mu = P^\mu_{\,\,\nu} \Psi^\nu, \\
    \partial_0 X^\mu = i P^\mu_{\,\,\nu} \partial_0 X^\nu +
    \frac{1}{4} Q^\mu_{\,\,\nu} N^\nu_{\,\,\sigma\rho}
    \Psi^\sigma \Psi^\rho , \\
    \widetilde{\Psi}^\mu = Q^\mu_{\,\,\nu} \widetilde{\Psi}^\nu, \\
    f^\mu = Q^\mu_{\,\,\nu} f^\nu +\frac{1}{4} Q^\mu_{\,\,\nu}
    N^\nu_{\,\,\sigma\rho} \Psi^\sigma \widetilde{\Psi}^\rho ,
\end{array} \right.
\eeq{e:1Dbc}
where $N^\nu_{\,\,\sigma\rho}$ is the Nijenhuis tensor for $R^\mu_{\,\,\nu}$,
(\ref{Ntdef2}).
In terms of the 1D superfields, these may be concisely written as
\begin{eqnarray}
\nonumber DK^\mu &=& P^\mu_{\,\,\nu} (K) DK^\nu, \\
 S^\mu &=& Q^\mu_{\,\,\nu}(K) S^\nu .
\label{e:1Dsuperbc}
\end{eqnarray}
It is clear from conditions (\ref{e:1Dsuperbc}) that the multiplet
$(X^\mu, \Psi^\mu)$ may be thought of as living along the Neumann
directions, whereas the multiplet $(\tilde{\Psi}^\mu, f^\mu)$ lives in
the Dirichlet directions.

\section{Supercurrents}
\label{a:supercurrents}

Here we briefly sketch how to derive the supercurrents that define the
supersymmetry currents and stress tensor discussed in
Section~\ref{s:currents}. Our derivation here includes a nonvanishing
background $B$-field; to obtain the currents relevant to the $B=0$
case, one just puts $B_{\mu\nu,\rho}=B_{\mu\nu}=0$ in the result
below.

We start from the superspace formulation of the non-linear sigma
model, where we have promoted the worldsheet to a superspace by
supplementing the ordinary worldsheet coordinates $\xi^\pm$ with a
pair of Grassmann coordinates $\theta^\pm$. The model has the same
form as (\ref{acrbas}), except now we make it locally supersymmetric by
introducing the supervielbein $E_M^{\,\,\,A}$ as well as replacing the
flat superderivatives $D_\pm$ by covariant ones, $\nabla_\pm$.  We
have
\beq
S=  \int d^2\xi d^2\theta\,\, E
      \nabla_{+} \Phi^\mu \nabla_{-} \Phi^\nu e_{\mu\nu} (\Phi),
\eeq{e:Sviel}
where $E$ is the determinant of the supervielbein.  The indices run
over the lightcone coordinates and their Grassmann counterparts, $M$
and $A$ taking values $(\+, =, +, -)$. The superfield $\Phi^\mu$ is
defined in Appendix~\ref{a:11susy}, and $e_{\mu\nu}$ is the superfield
whose lowest component is the spacetime metric plus B-field,
$e_{\mu\nu}=g_{\mu\nu}+B_{\mu\nu}$.

Varying (\ref{e:Sviel}) with respect to the independent
components of the supervielbein, we
obtain an expression of the form
\beq
 \delta S= \half \int d^2\xi d^2\theta\,\, E \sum_A (-1)^A
   T_A^{\,\,\,B} H_B^{\,\,\,A},
\eeq{e:TAB}
where $T_A^{\,\,\,B}$ are the supercurrents. We take the independent
variations to be ($H_+^{\,\,\,+}$, $H_-^{\,\,\,-}$,
$H_\pm^{\,\,\,\+}$, $H_\pm^{\,\,\,=}$), where $H_B^{\,\,\,A} \equiv
\delta E_A^{\,\,\,M} E_M^{\,\,\,B}$ \cite{Bastianelli}. Using the
equations of motion,
$$
\nabla_+ \nabla_- \Phi^\nu g_{\mu\nu} -
\half \nabla_+ \Phi^\rho \nabla_- \Phi^\nu
\left( g_{\mu\rho,\nu} + g_{\rho \nu,\mu} - g_{\mu\nu,\rho}
+  B_{\mu\nu,\rho} + B_{\rho\mu,\nu} + B_{\nu\rho,\mu} \right)
=0,
$$
as well as their $\nabla_\pm$ derivatives, all components
$T_A^{\,\,\,B}$ vanish except $T_{\+}^-$ and $T_{=}^+$.  To revert to
the case with global supersymmetry, we reduce the covariant
derivatives to flat ones again, and we get
\ber
&& T_{\+}^- = D_{+}\Phi^\mu \partial_{\+} \Phi^\nu g_{\mu\nu} -
\frac{i}{2}
D_{+}\Phi^\mu D_{+} \Phi^\nu D_{+} \Phi^\rho B_{\mu\nu,\rho} \\
&& T_{=}^+ = D_{-}\Phi^\mu \partial_{=} \Phi^\nu g_{\mu\nu} +
\frac{i}{2} D_{-}\Phi^\mu D_{-} \Phi^\nu D_{-} \Phi^\rho
B_{\mu\nu,\rho} \eer{e:Tcomps}
These are the supercurrents that define the supersymmetry current and
stress tensor via Eqns~(\ref{comp1})--(\ref{TTdefinition}).

\section{Almost product manifolds}
\label{a:APM}

Here we review the relevant mathematical definitions pertaining to
almost product manifolds.  In our use of terminology we closely follow
Yano's books \cite{Yano1,Yano2}; however, the reader should be aware
that often a different terminology is used in the literature.

Let ${\cal M}$ be a $d$-dimensional manifold with a
$(1,1)$ tensor $R^\mu_{\,\,\nu}$ such that, globally,
\beq 
 R^\mu_{\,\,\nu}R^\nu_{\,\,\rho}=\delta^\mu_{\,\,\rho} .
\eeq{defaps}
Then ${\cal M}$ is an almost product manifold with almost
product structure $R^\mu_{\,\,\nu}$.

If ${\cal M}$ admits a  Riemannian
metric $g_{\mu\nu}$ such that
\beq
 g_{\mu\nu} = R^\rho_{\,\,\mu}
  R^\sigma_{\,\,\nu} g_{\rho\sigma},
\eeq{apRman}
then ${\cal M}$ is an almost product Riemannian
manifold\footnote{This is not much of a restriction, since if the
  manifold allows a metric $\tilde{g}_{\mu\nu}$,
  then $g_{\mu\nu} = \tilde{g}_{\mu\nu}+R^\rho_{\,\,\mu}
  R^\sigma_{\,\,\nu} \tilde{g}_{\rho\sigma}$ will satisfy
  (\ref{apRman}).}.

One may define the following two tensors on ${\cal M}$,
\beq
 P^\mu_{\,\,\nu}=\frac{1}{2}(\delta^\mu_{\,\,\nu}+R^\mu_{\,\,\nu}),
\,\,\,\,\,\,\,\,\,\,\,\,\,\,\,\,\,\,\,\,\,
 Q^\mu_{\,\,\nu}=\frac{1}{2}(\delta^\mu_{\,\,\nu}- R^\mu_{\,\,\nu}).
\eeq{e:defPQ} 
$P$ and $Q$ are globally defined and
satisfy $R^\mu_{\,\,\nu} = P^\mu_{\,\,\nu}
- Q^\mu_{\,\,\nu}$. They are orthogonal projectors, i.e.,
$$
Q^\mu_{\,\,\nu}P^\nu_{\,\,\rho}=P^\mu_{\,\,\nu}Q^\nu_{\,\,\rho}=0
$$
and
$$
P^\mu_{\,\,\nu}P^\nu_{\,\,\rho}=P^\mu_{\,\,\rho},
\,\,\,\,\,\,\,\,\,\,\,\,\,\,\,
Q^\mu_{\,\,\nu}Q^\nu_{\,\,\rho}=Q^\mu_{\,\,\rho}.
$$

The eigenvalues of $R$ with respect to $P$ and $Q$ are $+1$ and $-1$,
respectively. Thus $P$ and $Q$ define two complementary distributions
such that tangent vectors $v^\mu$ on ${\cal M}$ with $R$-eigenvalue
$+1$ belong to $P$, and vectors with $R$-eigenvalue $-1$ belong to
$Q$,
$$
P: \,\,\,\,\, \{v^\mu : R^\mu_{\,\,\nu}v^\nu = v^\mu \} ,
\,\,\,\,\,\,\,\,\,\,\,\,\,\,\,
Q: \,\,\,\,\, \{v^\mu : R^\mu_{\,\,\nu}v^\nu = -v^\mu \} .
$$
Clearly, $\rank(P)$ equals the number of $+1$ eigenvalues of $R$, and
$\rank(Q)$ is the number of $-1$ eigenvalues of $R$. In fact,
this implies that the holonomy
group of an almost product Riemannian manifold is\footnote{For a
  \emph{pseudo} Riemannian manifold the holonomy group is
  $O(1,\rank(P)-1) \times O(\rank(Q))$.}  $O(\rank(P)) \times
O(\rank(Q))$.

At any one point $x^\mu_0$ in ${\cal M}$, it is always possible to find a
local coordinate basis such that $R$, $P$ and $Q$ take their canonical
form
\beq
R^\mu_{\,\,\nu}=
\left( \begin{array}{cc}
    \delta^m_{\,\,n} &  0 \\
    0 & -\delta^i_{\,\,j}
\end{array} \right),
\,\,\,\,\,\,\,\,\,
P^\mu_{\,\,\nu}=
\left( \begin{array}{cc}
    \delta^m_{\,\,n} &  0 \\
    0 & 0
\end{array} \right),
\,\,\,\,\,\,\,\,\,
Q^\mu_{\,\,\nu}=
\left( \begin{array}{cc}
    0 &  0 \\
    0 & \delta^i_{\,\,j}
\end{array} \right).
\eeq{e:RPQcan}

The almost product structure $R$ as well as the distributions $P$ and
$Q$ may or may not be \emph{integrable}.  The integrability condition
for $P$ and $Q$ are stated as follows.\footnote{Integrability may also
  be expressed in terms of the Frobenius theorem.} $P$ is completely
integrable if
\beq
P^\mu_{\,\,\gamma} P^\nu_{\,\,\sigma} Q^\rho_{\,\,[\mu,\nu]} =
\frac{1}{8}(N^\rho_{\,\,\gamma\sigma} - R^\rho_{\,\,\mu}
N^\mu_{\,\,\gamma\sigma}) = \frac{1}{4}
Q^\rho_{\,\,\mu}N^\mu_{\,\,\gamma\sigma}=0
\eeq{integrablP}
where $N^\mu_{\,\,\gamma\sigma}$ is the Nijenhuis tensor for $R$,
\beq
N^\rho_{\,\,\mu\nu} = R^\gamma_{\,\,\mu} R^\rho_{\,\,[\nu,\gamma]} -
R^\gamma_{\,\,\nu} R^\rho_{\,\,[\mu,\gamma]}.
\eeq{Ntdef2}
$Q$ is completely integrable if
\beq
Q^\mu_{\,\,\gamma} Q^\nu_{\,\,\sigma} P^\rho_{\,\,[\mu,\nu]} =
\frac{1}{8}(N^\rho_{\,\,\gamma\sigma} + R^\rho_{\,\,\mu}
N^\mu_{\,\,\gamma\sigma}) = \frac{1}{4}
P^\rho_{\,\,\mu}N^\mu_{\,\,\gamma\sigma}=0 .
\eeq{integrablP11}
$R$ is integrable if both $Q$ and $P$ are integrable, i.e., if
$N^\rho_{\,\,\mu\nu} =0$. In this case ${\cal M}$ is an integrable
almost product manifold, also called a locally product manifold.

Integrability determines the extent to which $R$, $P$ and $Q$ may keep
their canonical form in a local neighbourhood of the point $x^\mu_0$.
On an integrable almost product manifold, $R$, $P$ and $Q$ can always
be brought to the form (\ref{e:RPQcan}) in a whole neighbourhood of
$x^\mu_0$.  However, if only one of $P$ and $Q$ is integrable, then
the canonical form can be extended only in the corresponding
directions. Thus, if $P$ is integrable, one can adopt (\ref{e:RPQcan})
along the $P$-directions, and similarly for $Q$-integrability.

In terms of transition functions on ${\cal M}$, $R$-integrability
means that there is a system of coordinate neighbourhoods with
coordinates $X^\mu$ splitting into $(X^n, X^i)$ such that the
transition functions $f^\mu$ are of the form $\tilde{X}^n=f^n(X^m)$
and $\tilde{X}^i = f^i(X^j)$ so that $f^n_{\,\, ,i}=0$ and $f^i_{\,\,
  ,n}=0$. This is analogous to the case of almost complex manifolds
(where $R^2=-1$); there $R$-integrability implies that one can find
local (anti)holomorphic coordinates with (anti)holomorphic transition
functions.

If, for a locally product manifold ${\cal M}$, $R^\mu_{\,\,\nu}$ is a
covariantly constant tensor (i.e., $\nabla_\rho R^\mu_{\,\,\nu}=0$
with respect to some connection), then the manifold is called a
locally decomposable  Riemannian manifold. The warped product
manifold (see discussion at the end of Section~\ref{s:geometric}) is
an example of a locally product manifold which is not locally
decomposable.

\end{document}